# All-optical switching in rubidium vapor

Andrew M. C. Dawes, Lucas Illing, Susan M. Clark, and Daniel J. Gauthier*

*Duke University, Department of Physics, Box 90305, Durham, North Carolina 27708, USA*

**Summary Sentence**: We realize an optical switch that operates in the quantum regime with a weak beam controlling the direction of much stronger beams.

* **Corresponding author**:  Daniel J. Gauthier
　　　　　　　　　　　　　　Duke University
　　　　　　　　　　　　　　Department of Physics
　　　　　　　　　　　　　　Box 90305
　　　　　　　　　　　　　　Durham, NC 27708  USA
　　　　　　　　　　　　　　e-mail: gauthier@phy.duke.edu
　　　　　　　　　　　　　　Voice: (919) 660-2511
　　　　　　　　　　　　　　FAX: (919) 660-2525

**Abstract**

We report on an all-optical switch that operates at low light levels. It consists of laser beams counterpropagating through a warm rubidium vapor that induce an off-axis optical pattern. A switching laser beam causes this pattern to rotate even when the power in the switching beam is much lower than the power in the pattern. The observed switching energy density is very low, suggesting that the switch might operate at the single-photon level with system optimization. This approach opens the possibility of realizing a single-photon switch for quantum information networks and for improving transparent optical telecommunication networks.

An important component of high-speed optical communication networks is a switch capable of redirecting pulses of light *(1)*, where an incoming "switching" beam redirects other beams via light-by-light scattering in a nonlinear optical material *(2)*. For quantum information networks, it is important to develop optical switches that are actuated by a single photon *(3)*. Unfortunately, the nonlinear optical interaction strength of most materials is so small that achieving single-photon switching is exceedingly difficult. This problem appears to be solved through modern quantum-interference methods, where the nonlinear interaction strength can be increased by many orders-of-magnitude *(3-10)*. Another desirable property of all-optical switches is that the output beams are controlled by a weaker switching beam so they can be used as cascaded classical or quantum computational elements *(11)*. Current switches, however, tend to control a weak beam with a strong one.

In this Report, we describe an all-optical switch that combines the extreme sensitivity of instability-generated transverse optical patterns to tiny perturbations *(12-16)* with quantum interference methods *(3-10)*. A transverse optical pattern is the spatial structure of the electromagnetic field in the plane perpendicular to the propagation direction. As an example, the transverse optical pattern corresponding to two beams of light is a pair of spots. We control such a pattern with a beam whose power is up to 6,500 times weaker than the power contained in the pattern itself, verifying that the switch is cascadable. Also, the switch is actuated with as few as 2,700 photons and thus operates in the low-light-level regime. A measured switching energy density $\mathcal{E} \sim 3 \times 10^{-3}$ photons/$(\lambda^2/2\pi)$, where $\lambda = 780$ nm is the wavelength of the switching beam, suggests that the switch might operate at the single-photon level with system optimization such as changing the pump-beam size or vapor cell geometry *(17)*.

Our experimental setup consists of a weak switching beam that controls the direction of laser beams emerging from a warm laser-pumped rubidium vapor. Two pump laser beams counterpropagate through the vapor and induce an instability that generates new beams of light (*i.e.* a transverse optical pattern) when the power of the pump beams is above a critical level *(17)* as shown schematically in Fig. 1. The instability arises from mirror-less parametric self-oscillation *(17-26)* due to the strong nonlinear coupling between the laser beams and atoms. Mirror-less self-oscillation occurs when the parametric gain due to nonlinear wave-mixing processes becomes infinite. Under this condition, infinitesimal fluctuations in the electromagnetic field strength trigger the generation of new beams of light. The threshold for this instability is lowest (and the parametric gain enhanced) when the frequency of the pump beams is set near the $^{87}$Rb $5S_{1/2} \leftrightarrow 5P_{3/2}$ resonance (780-nm transition wavelength). The setup is extremely simple in comparison to most other low-light-level all-optical switching methods *(7, 8)* and the spectral characteristics of the switching and output light match well with recently demonstrated single-photon sources and storage media *(27, 28)*.

For a perfectly symmetric experimental setup, the instability-generated light (referred to henceforth as "output" light) is emitted both forward and backward along cones centered on the pump beams, as shown in Fig. 1A. The angle between the pump-beam axis and the cone is of the order of ~5 mrad and is determined by competition between two different nonlinear optical processes: backward four-wave mixing in the phase-conjugation geometry and forward four-wave mixing *(17, 23, 24)*. The generated light has a state-of-polarization that is linear and orthogonal to that of the linearly co-polarized pump beams *(25)*; hence, it is easy to separate the output and pump light using polarizing elements. Once separated, the output light propagating in one direction (say, the forward direction) can appear as a ring on a measurement screen that is perpendicular to the propagation direction and in the far field (Figs. 1A and 1B). This ring is known as a transverse optical pattern *(18)* and is one of many patterns that occur in a wide variety of nonlinear systems spanning the scientific disciplines *(29)*.

Weak symmetry breaking caused by slight imperfections in the experimental setup reduces the symmetry of the optical pattern and selects its orientation *(23)*. For high pump powers, the pattern consists of 6 spots with 6-fold symmetry, as shown schematically in Fig. 1C. For lower powers near the instability threshold, only two spots appear in the far field in both directions (forward and backward), as shown schematically in Fig. 1D. The azimuthal angle of the spots (and the corresponding beams) and is dictated by the system asymmetry, which can be adjusted by slight misalignment of the pump beams or application of a weak magnetic field. The orientation of the spots is stable for several minutes in the absence of a switching beam.

In our all-optical switch, the direction of the bright output beams (total power $P_{out}$) is controlled by applying a weak switching laser beam whose state of polarization is linear and orthogonal to that of the pump-beams (Fig. 2.) The azimuthal angle of the output beams is extremely sensitive to tiny perturbations because the symmetry breaking of our setup is so small *(17)*. Directing the switching beam along the conical surface at a different azimuthal angle (see Fig. 2B) causes the output beams to rotate to a new angle, while $P_{out}$ remains essentially unchanged. Typically, the orientation of the output beams rotates to the direction of the switching beam and we find that the pattern is most easily perturbed when the switching beam is injected at azimuthal angles of ±60°, thereby preserving the 6-fold symmetry of the pattern observed for higher pump powers. It should be noted that the switching beam also controls the output beams in the backward direction.

We now describe the behavior of our switch for two values of the peak power $P_s$ of the switching beam, where the spots rotate by -60° in the presence of a switching beam. In the absence of a switching beam ($P_s = 0$), we observe the pattern shown in Fig. 3A, where $P_{out} = 1.5$ µW and the total power emitted from the vapor cell in the forward direction in both states of polarization is 19 µW. For the higher power switching beam ($P_s = 2.5$ nW) we observe complete rotation of the output beams (Fig. 3B), whereas we find that only approximately half the power in the beams rotates to the new azimuthal angle (Fig. 3C) at the lower switching power ($P_s = 230$ pW). However, we observe high-contrast switching in both cases.

To quantify the dynamic behavior of the device, we turn the switching beam on-and-off and measure the resulting change in the output beams. Figures 4A and B show the temporal evolution of the power for $P_s = 2.5$ nW on a coarse time, where it is seen that power variation at each location is out of phase, indicating full redirection of the optical power. (See also Movie S1.)

At higher temporal resolution (Figs. 4C), weak periodic modulation of the emitted light due to a dynamic instability *(25)* is apparent. This modulation corresponds to small fluctuations in the power of the pattern, although its orientation is stable. Even in the presence of this modulation, the contrast-to-noise-ratio of the switch is at least 30:1 (defined as the change in power between the on and off states : root-mean-square value of the fluctuations). The rise-time of the switch is $\tau = 4$ μs, which we believe is largely governed by the rubidium ground-state optical pumping time *(23)*.

Under the conditions shown in Figure 3B, the total power in the output beams in the forward direction is equal to $P_{out} = 1.5$ μW, while the power of the input switching beam is only $P_s = 2.5$ nW. That is, a switching beam controls the behavior of output beams that are at least 600 times more powerful. Based on the time response of the switch, the number of photons needed to change its state is given by $N_p = \tau P_s /E_p =$ 40,000, where $E_p = 2.55 \times 10^{-19}$ J is the photon energy, and the switching energy is equal to $N_p E_p = 10$ femtoJ. Optical switches that operate with similarly low energies have been proposed previously, but their application to few-photon switching has not been discussed *(12, 14, 15)*.

Another metric to characterize low-light-level switches is the energy density $\mathcal{E}$, given in units of photons per $(\lambda^2/2\pi)$. This metric gives the number of photons needed to actuate a switch whose transverse dimension has been reduced as small as possible - the diffraction limit of the interacting beams *(5, 11)*. For the spot size of the switching beam used in our experiment (1/e intensity radius of 166 μm), we find that $\mathcal{E} \sim 4.4 \times 10^{-2}$ photons/$(\lambda^2/2\pi)$, corresponding to 11 zeptoJ/$(\lambda^2/2\pi)$.

Similar behavior is observed for the lower switching power, as shown in Fig. 3C. In this case, a weak switching beam controls output beams that are 6,500 times stronger. Even though there is only partial switching, the contrast-to-noise ratio is greater than 10:1 (Figs. 4, D and E; see also Movie S2). As seen in Fig. 4F, $\tau = 3$ μs, which is somewhat faster than that observed at the higher power, possibly due to the fact that only part of the output light rotates to the new angle. Using this response time, we find $N_p = 2,700$ photons, $N_p E_p = 690$ attoJ, and $\mathcal{E} \sim 3 \times 10^{-3}$ photons/$(\lambda^2/2\pi)$ [770 yoctoJ/$(\lambda^2/2\pi)$].

These results demonstrate that a switch based on transverse optical patterns is capable of controlling high power beams with weak ones, exhibits much higher sensitivity, and can be realized using a simple experimental setup. Such a switch could be used as a binary element in a computation or communication system because, at high switching beam power, our device operates like a transistor used in digital logic where the transistor's output is either off or saturated. Additionally, this switch could be used as a router if information is impressed on the output light (*e.g.*, by modulating the pump beams).

For comparison, the sensitivity of our switch, characterized by $\mathcal{E}$, far exceeds that demonstrated using other methods such as electromagnetically-induced transparency (EIT). To date, the best EIT-based switch results have been reported by Braje *et al.* *(8)* who have achieved $\mathcal{E} \sim 23$ photons/$(\lambda^2/2\pi)$; our switch is over 5,000 times more sensitive. Furthermore, the EIT experimental setup is much more complicated, requiring the use of cooled and trapped rubidium atoms, and limited to output beams that are much weaker than the input switching beam. Recently, a transient switch based on laser beams propagating through a warm rubidium vapor in a simple setup has been reported, but it does not operate in the low-light-level regime *(10)*.

While speculative, our technique might be useful at telecommunication wavelengths where high-quantum efficiency, low-noise, single-photon detectors are difficult to realize. For these wavelengths, the rubidium vapor could be replaced with a molecular gas, such as acetylene or hydrogen cyanide, both of which have resonances in the telecommunications band. A low-light-level incident beam could then be detected by switching the output beam onto a standard telecommunications-band detector.

In addition, our general method of exploiting the sensitivity of patterns to small perturbations may find application in other scientific disciplines. For example, modulational instabilities, which often give rise to pattern formation, have been observed in matter waves created with ultra-cold quantum gases *(30)*, suggesting that atom switching might be possible by perturbing the gas with a few injected atoms.

Our results may also have implications concerning the fundamental limits of general-purpose computation devices. Many years ago, Keyes *(11)* realized that thermal energy dissipation places limits on the operational speed of a logic element. By assuming that a saturation-based optical switch has $E \sim 1$ photon/$(\lambda^2/2\pi)$, he concluded that optical logic elements are limited to rates below $10^{10}$ s$^{-1}$. Our observed switching energy density is over a factor of 300 below that assumed by Keys, suggesting that optical devices might surpass his estimated limit.

**Supporting Online Material**

www.sciencemag.org

Materials and Methods

Figure S1, S2

Movies S1, S2

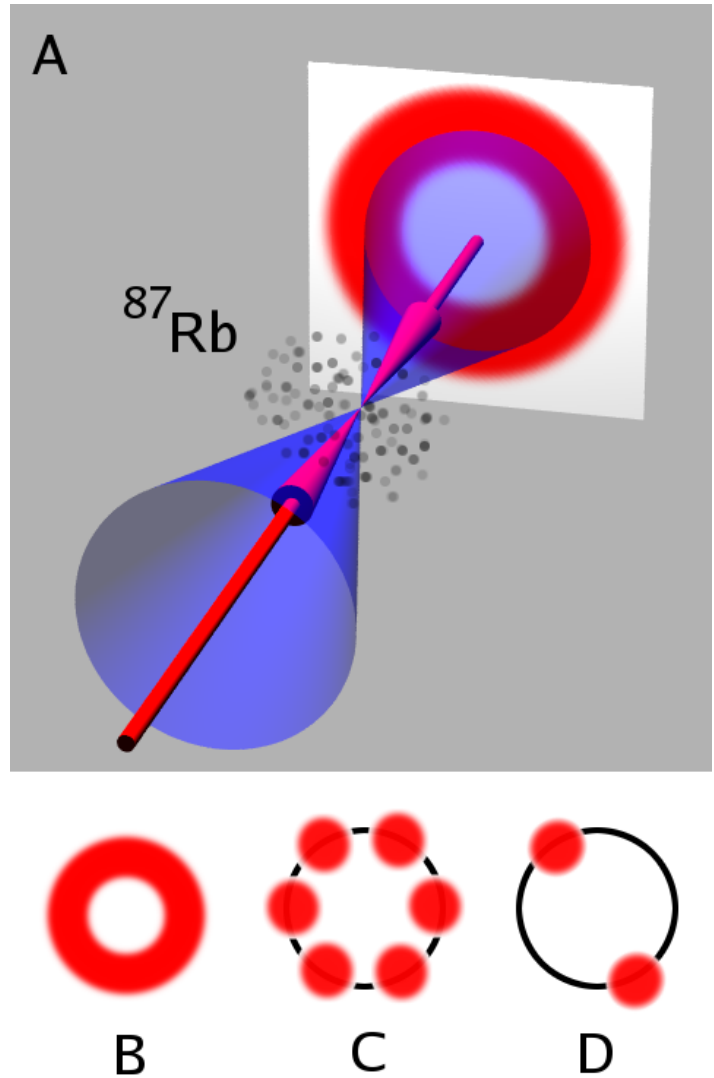

**Figure 1:** Generation and symmetry of transverse optical patterns. (**A**) Two linearly co-polarized pump beams (red) counterpropagate through warm $^{87}$Rb vapor. A modulational instability generates orthogonally polarized light, which is emitted along cones (blue) and forms a transverse optical pattern (red) on a screen perpendicular to the propagation direction. (**B-D**) Schematic of the transverse optical pattern: (**B**) for a perfectly symmetric setup, (**C**) for weakly broken symmetry and high pump power, and (**D**) for weakly broken symmetry and low pump power.

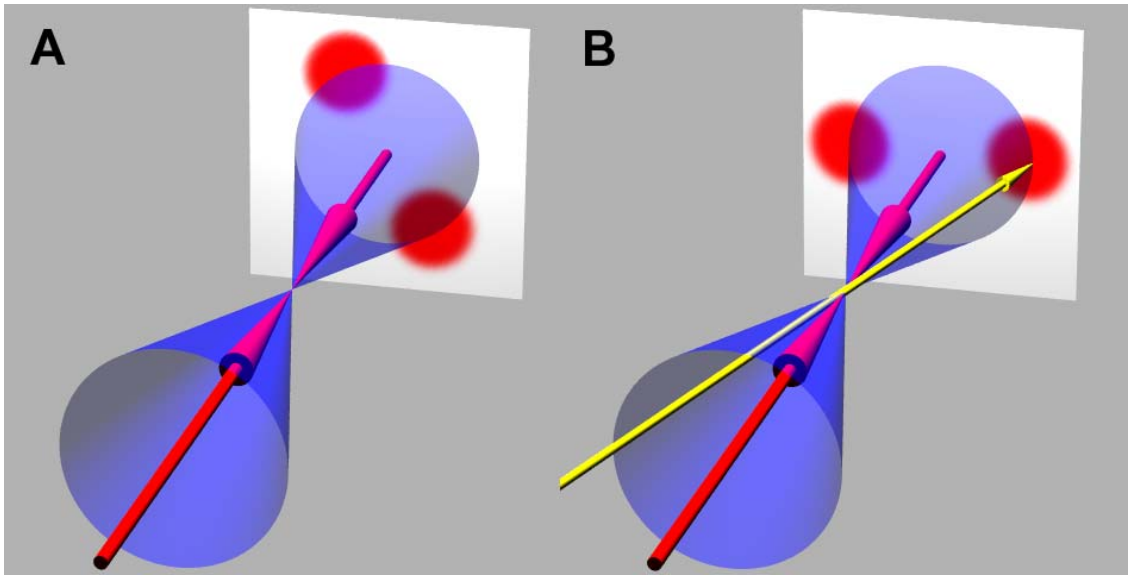

**Figure 2:** The two states of the switch. (**A**) The off state. Weak symmetry breaking results in a two-spot output pattern. (**B**) The on state. A weak switching beam (yellow), directed along the cone (blue), causes the output pattern to rotate. The state of polarization of the switching beam (yellow) is linear and orthogonal to that of the pump beams (red).

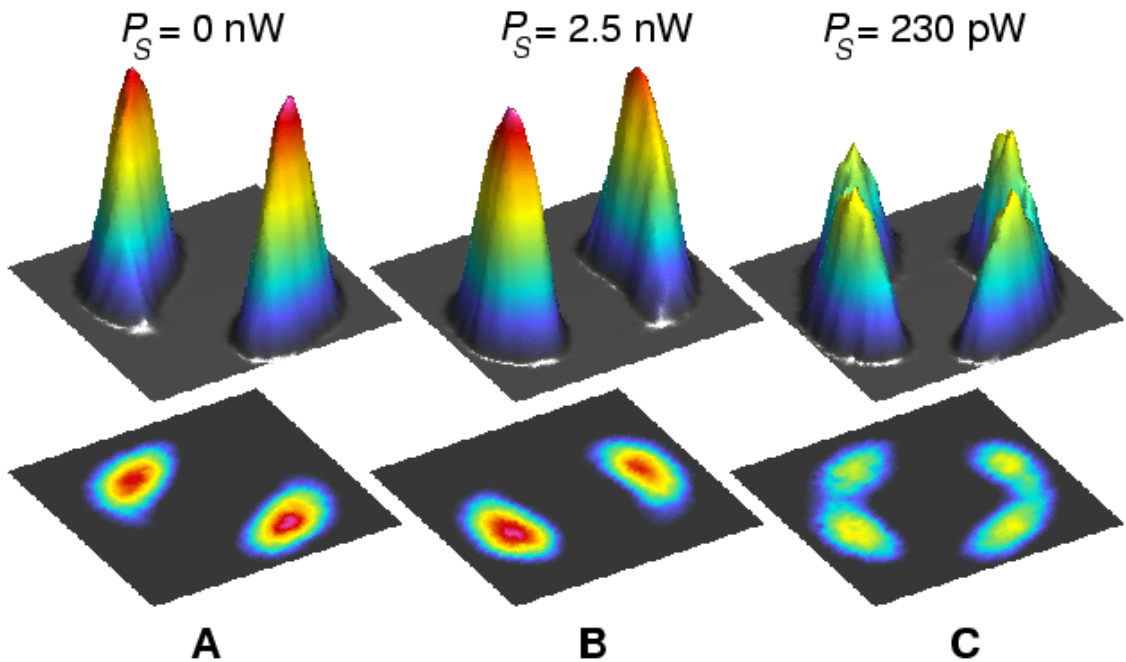

**Figure 3:** Low-light-level all-optical switching. The lower panels show a false-color rendition of the detected optical power of the output light (grey: low power, red: high power), and the upper panels are a three-dimensional representation of the same data. (**A**) The off state with $P_s = 0$. The output light forms a two-spot transverse optical pattern. (**B**) The on state with $P_s = 2.5$ nW. The two-spot output pattern is rotated by -60°. (**C**) The on state with $P_s = 230$ pW. Approximately half the output power is rotated by -60°.

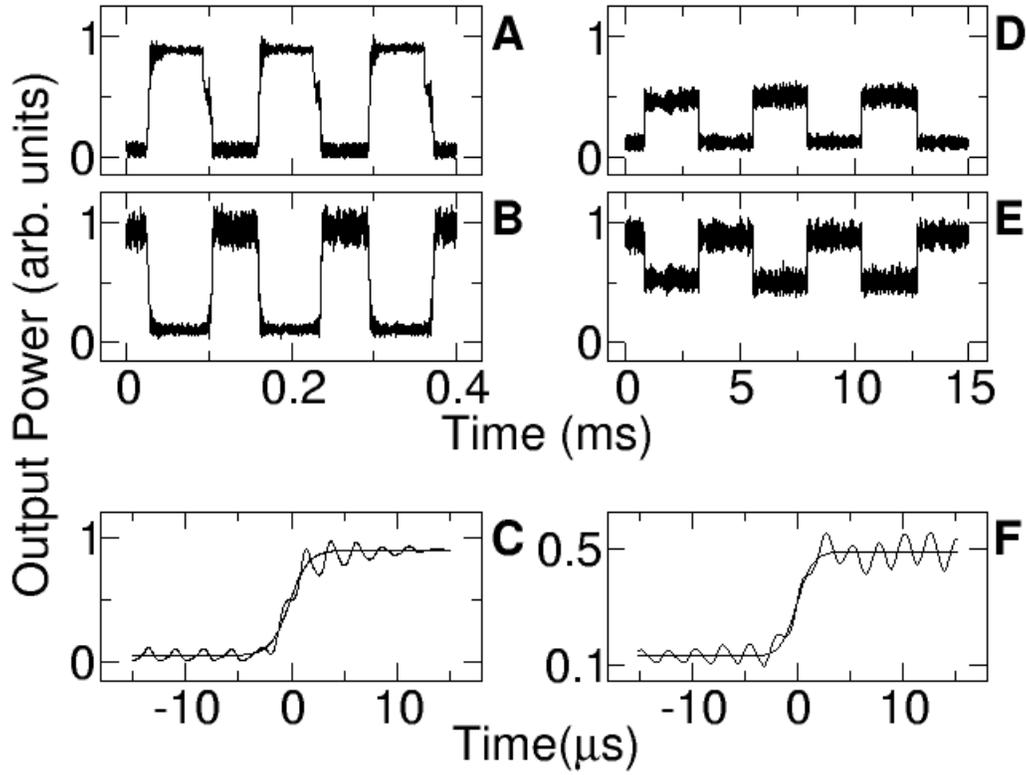

**Figure 4:** Dynamical behavior of the low-level all-optical switch. In the plane of the measurement screen, we place one aperture at the center of one of the spots shown in Fig. 3A (the "off state" of the switch), and another at the center of one of the spots shown in Fig. 3B (the "on state"). Temporal evolution of the output power passing through (**A** and **C**) the "on-state" aperture for $P_s = 2.5$ nW, (**B**) the "off-state" aperture for $P_s = 2.5$ nW, (**D** and **F**) the "on-state" aperture for $P_s = 230$ pW, and (**E**) the "off-state" aperture for $P_s = 230$ pW. The fit lines in (**C**) and (**F**) are determined using a sigmoidal function. The rise-time of the switch based on this fit is: (**C**) $\tau = 4$ μs for $P_s = 2.5$ nW and (**F**) $\tau = 3$ μs for $P_s = 230$ pW.

**Supporting Online Material**

**Materials and Methods**

The laser beams used to pump the $^{87}$Rb vapor are derived from the output of a frequency-stabilized continuous-wave Ti:sapphire laser, which is spatially filtered using a single-mode optical fiber with angled ends, and collimated to a spot size (1/e field radius) of 470 μm. The input power of one of the beams (denoted as the forward beam) has a power of 630 μW and the other (denoted as the backward beam) has a power of 250 μW. Polarizing beam splitters are placed in each beam so that the input beams are linearly polarized with parallel polarizations.

The isotopically-enriched rubidium vapor (>90% $^{87}$Rb) is contained in a 5-cm-long glass cell heated to 70 °C (corresponding to an atomic number density of ~7 × 10$^{11}$ atoms/cm$^3$). The cell is tilted with respect to the incident laser beams to prevent possible oscillation between the uncoated windows. The cell has no paraffin coating on the interior walls to prevent depolarization of the ground-state coherence, nor does it contain a buffer gas that would slow diffusion of atoms out of the pump laser beams. The Doppler-broadened linewidth of the transition at this temperature is ~550 MHz. To prevent the occurrence of magnetically-induced instabilities, we us a Helmholtz coil to cancel the ambient magnetic field component along the direction of the counterpropagating laser beams. We believe it is not necessary to use an isotopically-enriched rubidium vapor; it is used as a matter of convenience.

The linearly and co-polarized pump-laser beams counterpropagate through the $^{87}$Rb vapor and induce an instability that generates light in the orthogonal state of polarization (Fig. S1). The instability arises from mirror-less self-oscillation, which occurs when the parametric gain due to nonlinear wave-mixing processes becomes formally infinite. Physically, this implies that new beams of light are spontaneously generated and emitted from the vapor *(S1, S2)*.

The instability-generated output light is an emergent spatial structure in the plane orthogonal to the propagation direction. This structure is referred to as a transverse optical pattern *(S3)*. In general, the mechanism that gives rise to the pattern formation is the combination of nonlinearity and coupling between the different spatial points due to diffraction *(S3)*. Specifically, the pattern forming instability in our system can be interpreted in terms of either optical pumping or four-wave mixing processes *(S4)*.

Several experiments have investigated instabilities that occur when laser beams counterpropagate through an atomic vapor and demonstrated spontaneous formation of various transverse optical patterns, including rings and multi-spot off-axis patterns in agreement with our observations *(S4-S6)*. A complete theoretical treatment of our experimental system is complicated because the absorption and saturation of the nonlinearity due to optical pumping, the vector nature of the electric field, and the atomic motion are all potentially important. Although no full treatment of this system has been given in the existing literature, several theories predict the general features observed in our experiment, such as the emission of light along a cone and the formation of stable transverse optical patterns with 2- and 6-fold symmetry *(S7-S10)*.

In our switching experiment we use a two-spot transverse optical pattern that is stationary in the sense that the transverse spatial structure of the pattern is static, in the absence of a switching beam, over a period of many minutes. Drift of the pattern on this long time scale is primarily due to drift in the laser frequency or pump-beam alignment.

To characterize the instability in greater detail, we measure the total power of the output light as a function of the pump laser frequency. We observe several sub-Doppler features in the power emitted from the vapor cell due to the instability (in the state of polarization orthogonal to that of the input pump beams) as the laser frequency is scanned through the $^{87}$Rb $5S_{1/2} \leftrightarrow 5P_{3/2}$ transition (Fig. S2). The maximum power emitted in the orthogonal polarization occurs when the laser frequency ν is tuned ~25 MHz to the high-frequency side of the $5S_{1/2}$ (*F*=1) $\leftrightarrow$ $5P_{3/2}$ (*F''*=1) transition, where *F* (*F''*) denotes the total angular momentum of the hyperfine level of the ground (excited) state. We tune the laser frequency 5 MHz to the high-frequency side of this peak during the switching experiments.

We believe that it is possible to reduce the switching energy to the single-photon level through some combination of the following: adjusting the vapor length, reducing the size of both pump beams and the switching beam, adding a buffer gas, or using a vapor cell with paraffin-coated walls.

**Supporting References and Notes**

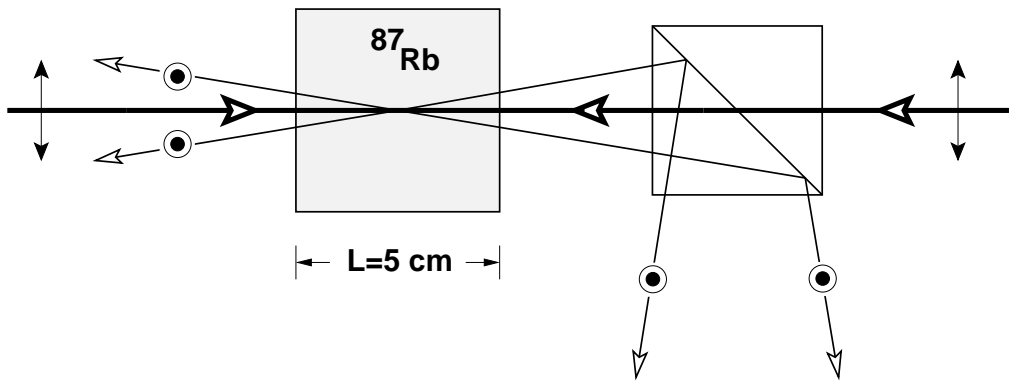

**Figure S1:** Mirror-less parametric self-oscillation generates new beams of light (*i.e.* a transverse optical pattern). Two pump beams (bold) counterpropagate through a 5 cm $^{87}$Rb vapor cell. The pump beams have identical linear polarization (shown as arrows in the plane of the page). The generated light is emitted off-axis in both forward and backward directions with a state of polarization that is linear and orthogonal (shown as circled dots) to that of the pump beams. The polarizing beam splitter (shown on the right) separates generated light from pump light.

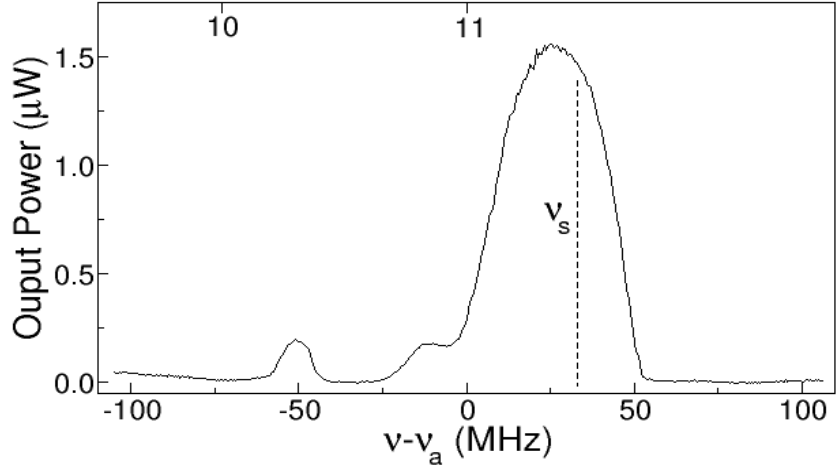

**Figure S2:** Resonant enhancement of the transverse optical pattern. Output power as function of the pump-beam frequency, where $\nu_a$ denotes the resonance frequency of the $5S_{1/2}$ ($F=1$) $\leftrightarrow$ $5P_{3/2}$ ($F'=1$) transition and $\nu_s$ denotes the frequency at which the switching data was collected. The top-axis tick marks "10" and "11" indicate the $5S_{1/2}$ ($F=1$) $\leftrightarrow$ $5P_{3/2}$ ($F'=0$) and the $5S_{1/2}$ ($F=1$) $\leftrightarrow$ $5P_{3/2}$ ($F'=1$) transition frequencies, respectively.

**Online Movies**

**Movie S1:** The low-level all-optical switch in action for a higher switching power. Temporal evolution of the emitted spots when the switching beam ($P_s = 2.5$ nW) is switched on-and-off at a rate of 1.6 s$^{-1}$.

**Movie S2:** The low-level all-optical switch in action for a lower switching power. Temporal evolution of the emitted spots when the switching beam ($P_s = 230$ pW) is switched on-and-off at a rate of 2.1 s$^{-1}$.